\documentclass[12pt,preprint]{aastex}

\newcommand{\msun}{\,\hbox{M$_{\odot}$}}
\newcommand{\kms}{\,\hbox{\hbox{km}\,\hbox{s}$^{-1}$}}

\usepackage{psfig}

\shorttitle{The Nuclear Gas Dynamics and Star Formation of Markarian\,231}
\shortauthors{}

\begin{document}

\title{The Nuclear Gas Dynamics and Star Formation of 
Markarian\,231\footnote{The near infrared data presented herein were
obtained at the W.M. Keck Observatory, which is operated as a
scientific partnership among the California Institute of Technology,
the University of California and the National Aeronautics and Space
Administration. The Observatory was made possible by the generous
financial support of the W.M. Keck Foundation.
}}

\author{R.I. Davies, L.J. Tacconi, and R. Genzel}
\affil{Max-Planck-Institut f\"ur extraterrestrische Physik, 
Postfach 1312, 85741, Garching, Germany}

\begin{abstract}
We report adaptive optics H- and K-band spectroscopy
of the inner few arcseconds of the luminous merger/ULIRG/QSO Mkn\,231,
at spatial resolutions as small as 0.085\arcsec.
For the first time we have been able to resolve the active star
forming region close to the AGN using stellar absorption features,
finding that its luminosity 
profile is well represented by an exponential function with a disk scale
length 0.18--0.24\arcsec\ (150--200\,pc), and implying that the stars
exist in a disk rather than a spheroid.
The stars in this region are also young (10--100\,Myr), and it therefore
seems likely that they have formed {\em in situ} in the gas disk, which
itself resulted from the merger.
The value of the stellar velocity dispersion ($\sim100$\kms\ rather
than the usual few times 10\kms\ in large scale disks) is a result of
the large mass surface density of the disk.
The stars in this region have a combined
mass of at least $1.6\times10^9$\msun, and account for
25--40\% of the bolometric luminosity of the entire galaxy.
At our spatial resolution the stellar light in the core is diluted by
more than a factor of 10 even in the H-band by
continuum emission from hot dust around the AGN.
We have detected the 2.12\micron\ 1-0\,S(1) H$_2$ and 1.64\micron\
[Fe{\sc ii}] lines out to radii exceeding 0.5\arcsec.
The kinematics for the two lines are very similar to each other as
well as to the stellar kinematics, and
broadly consistent with the nearly face-on rotating disk reported in
the literature and based on interferometric CO\,1-0 and CO\,2-1
measurements of the cold gas.
However, they suggest a more complex situation in which the inner
0.2--0.3\arcsec\ (200\,pc) is warped out of its original disk plane.
Such a scenario is supported by the projected shape of the nuclear
stellar disk, the major axis of which is significantly offset from the
nominal direction; 
and by the pronounced shift on very small scales in the direction of
the radio jet axis which has been reported in the literature.
\end{abstract}

\keywords{
galaxies: individual (Mkn\,231) ---
galaxies: Seyfert ---
galaxies: nuclei ---
galaxies: ISM ---
ISM: kinematics and dynamics ---
galaxies: star clusters}

\section{Introduction}
\label{sec:intro}

The ultraluminous infrared galaxy Mkn\,231 (UGC\,8058,
VII\,Zw\,490) has the highest luminosity 
($L_{\rm IR} \sim 3.2\times10^{12}$\,$L_\odot$ for a distance of
170\,Mpc) of the original ULIRG 
sample of \cite{san88} from the Bright Galaxy Survey.
That this galaxy hosts a powerful AGN is beyond doubt, as evidenced by
the wealth of observations:
for example, a compact nuclear radio continuum source with jets
\citep{nef88,ulv99a,ulv99b,tay99}, 
broad absorption lines \citep{smi95}, and a
compact source of variable hard X-ray emission \citep{gal02,bra03}.
Because Mkn\,231 has an absolute blue magnitude $M_{\rm B} = -21.7$
close to the cutoff of $-22.1$ for QSOs \citep{sch83}, it is often
classified as one of these.
However, it is also sometimes classed as a Seyfert\,1, particularly
because it has become clear that the AGN is responsible for only a
part of the luminosity, and that a significant fraction (1/3 to 2/3)
infact originates in star formation \citep{kra97,dow98,tac02}.
It is exactly this quality -- the co-existence in the nucleus of
both the star 
forming and AGN phenomena with massive luminosities -- combined with
its relative proximity that make
Mkn\,231 a key object for investigations of whether, as a
class, ULIRGs do \citep[e.g.][]{san88,vei02} or do not
\citep[e.g.][]{tac02} evolve into QSOs.

The asymmetrical optical morphology and large scale tidal tales
\citep{hut87,san87,can00} 
of Mkn\,231 are typical of the remnants predicted by models of merger
events between equal mass gas rich spiral galaxies
\citep[e.g.][]{bar92,mih99}.
Such models indicate that Mkn\,231 is most likely to be such a late
stage merger.
The structural and kinematic properties of a sample of similar late
stage ULIRG mergers have been studied by \cite{gen01} to investigate 
whether they might evolve later into (intermediate mass, $L_*$)
ellipticals;
and by \cite{tac02} to test whether, once rid of their gas and dust
shells, they might be the progenitors of QSOs.
The total sample consists of 18 ULIRGs, of which 8 contain
QSO nuclei.
Mkn\,231 presents a puzzle since its stellar velocity dispersion of
120\kms\ is the 
smallest of the sample (measured in 0.6\arcsec\ slits). 
A remnant of a merger between massive gas rich spirals is expected to have a
higher velocity dispersion, and 
indeed the ULIRG sample as a whole had a (logarithmic) average
of 185\kms.
On the other hand, Mkn\,231 appears to be similar to the rest of the
sample in having a relaxed stellar
population, with $V_{\rm rot}/\sigma=0.2$ and an $r^{1/4}$ luminosity
profile.
However, again the effective radius of the extended H and K band light
in MKn\,231 of 300\,pc is the smallest of the
sample, which had a (logarithmic) average of 2\,kpc.
Under the assumption that the stellar distribution is spheroidal,
\cite{tac02} calculated the mass within the effective radius (300\,pc)
to be $6\times10^9$\msun.
From the relation between velocity dispersion and black hole mass,
they also estimated $M_{\rm BH}=1.3\times10^7$\msun,
which would imply an Eddington efficiency of 6.
The rather uncomfortable result of a super-Eddington black hole
luminosity led them to consider whether the
stars might actually lie in a nearly face-on regularly rotating inner
disk, as exhibited by the cold molecular gas \citep{bry96,dow98},
since this could also lead to a low value for $V_{\rm rot}/\sigma$.
To address this question in detail and shed light on the nucleus of
Mkn\,231, we have undertaken near infrared spectroscopy using adaptive
optics.
The aim is to investigate its stellar luminosity profile, mass, and
geometry, and to probe its nuclear dynamics at high angular resolution
in the central arcsecond.

The paper is arranged as follows.
After summarising the observations in \S~\ref{sec:obs}, we consider
the shape and extent of the continuum emission in \S~\ref{sec:im}.
The radial profile, luminosity, and mass of the nuclear star forming
region is discussed in detail in \S~\ref{sec:starform}.
In \S~\ref{sec:kin} we turn to the kinematics of the gas and stars,
and construct a simple planar disk model which can reproduce their
characteristics.
Bringing these results together suggests that the starburst in the
nucleus of Mkn\,231 is very young, and so in \S~\ref{sec:young} we
appraise other evidence from the literature that this is the case.
Finally, we conclude in \S~\ref{sec:conc}.

\section{Observations and Data Reduction}
\label{sec:obs}

The data were obtained on the night of 28 May 2002 using NIRC-2, in a
partially commissioned shared risk basis,
behind the adaptive optics (AO) system \citep{wiz00} on the Keck II
telescope.
NIRC-2 is designed to take full advantage of the adaptive optics
system and has a $1024^2$ Aladdin-3 InSb array, with a number of grisms and
slits for the spectroscopic mode.
It is unusual in that, in order to select the required wavelength
range, the slit and science target are shifted in the AO system's
focal plane rather than tilting the grism.

\begin{figure}
\centerline{\psfig{file=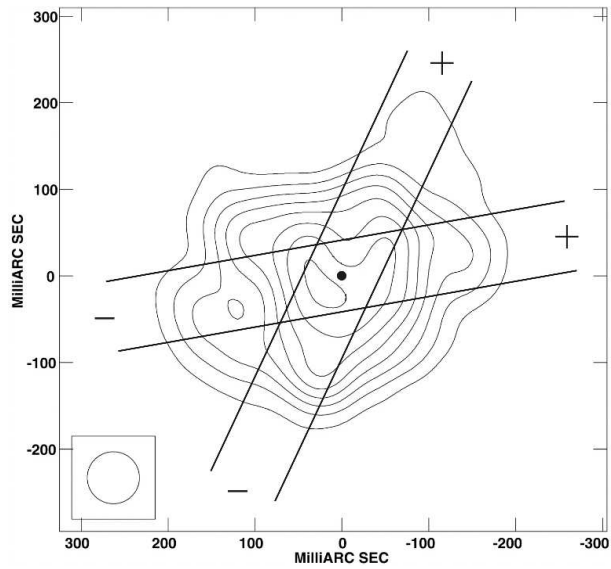,width=8cm}}
\caption{The two 0.08\arcsec\ slits at position angles $-80^\circ$ and
$-25^\circ$ superimposed on the 60\,mas resolution 1.4\,GHz radio
continuum map of \cite{car98} (taken from the Astronomical Journal).
The positive and negative ends of each slit (as referred to in the
text) have been identified as such.
It has been assumed that the slits were centered exactly on the
position of the central radio continuum source (marked by a filled
circle). 
}
\label{fig:slits}
\end{figure}

Spectra were obtained in the H- and K-bands with the camera in
widefield mode (40\,mas pixels) through a 80\,mas slit with the medium
resolution grism, resulting in nominal spectral resolutions of
$R=2385$ and 2090 respectively. 
The wavelength coverage is sufficient to include each band entirely in
a single setting.
Two position angles (PA) were used:
$-80^\circ$ (close to the kinematic major axis) and
$-25^\circ$ (reasonably close to the minor axis, but bisecting the
extended 1-0\,S(1) emission seen by \citealt{kra97}).
The two positions of the slit are depicted in Fig.~\ref{fig:slits}.
Individual exposures of 600\,sec in the H-band and 300\,sec in
the K-band were used, nodding the slit after each exposure.
Total integration times were 40\,mins and 80\,mins.
Standard calibrations were performed, including atmospheric standard
stars (using the type G2\,V star HD\,117845 in the H-band, and the F2
star HD\,110105 in the K-band), arcs, flatfields, and dark frames.
The data were reduced using PC-IRAF\,2.11.3 using standard techniques.

The spatial resolution depended on the adaptive optics performance,
the reference for which was Mkn\,231 itself.
Because the seeing was reasonably good and stable, the wavefront
sensor was able to run at a frame rate of 55\,Hz with up to 60\,counts
per pixel.
For the K-band this resulted in a spatial resolution limited only by
the coarse pixel sampling.
The FWHM measured on the nucleus of Mkn\,231 was 85\,mas (close to the
80\,mas expected for Nyquist sampling).
For the H-band, both the spatial and temporal scales of the
atmospheric turbulence are smaller, and hence the AO performance is
not so good.
The resulting FWHMs measured on the galaxy nucleus were 0.192\arcsec\
and 0.160\arcsec\ for the two position angles ($-80^\circ$ and
$-25^\circ$ respectively).
In Section~\ref{sec:starform} we show that the resolved stellar
continuum contributes relatively little to this, and hence we can take
it as a reasonable estimate of the spatial resolution.

The resulting FWHM spectral resolution measured from the arc lamp lines
is 120\kms\ at 2.2\micron, and 164\kms\ at 1.6\micron.
The K-band resolution is rather better than the nominal value because
of the AO correction which resulted in a PSF which had an intrinsic
FWHM rather less than 2 pixels across.
Hence the spectral resolution is limited by the PSF rather than simply
the slit width.

Flux calibration for the H-band data could not be performed using a
standard star. 
Instead we use the H-band point source magnitude for Mkn\,231 of
H=10.09 from 2MASS.
Extrapolating the continuum profile, assuming azimuthal symmetry, over
a filled circular aperture 
with a diameter of 2\arcsec\ (which contains effectively all the flux
we have detected; see Fig.~\ref{fig:profiles}) indicates that the flux
falling into the slit is a factor of 5.1 less than would be in such an
aperture.
Therefore we assign a magnitude of 11.86 to the flux detected in the
central 2\arcsec\ along the slit.
Similarly, the magnitude for a 1\arcsec\ circular aperture is H=10.19.


\section{Imaging \& Continuum Profile}
\label{sec:im}

A short sequence of images, totalling 20\,sec, were taken in the
K-band to assess the size of the nucleus of Mkn\,231 during adaptive
optics wavefront correction.
This was measured to be 85\,mas FWHM, including loss of resolution from
undersampling and sub-pixel shifting to align the frames.
A similar sequence of images taken on the standard star, 
while the AO system was running with the same control parameters
(although with 200\,counts rather than 60, even after dimming the star
with a neutral density filter), led to the same measurement of the
FWHM and an estimate of the Strehl ratio to be $\sim20$\% in the
K-band.
Thus, although the final images are not diffraction limited, this is
purely a result of the data acquisition procedure, and the nucleus of
Mkn\,231 is not resolved in the K-band on this spatial scale.

A second source about 0.19\arcsec\ to the north has been reported by
\cite{lai98} who claimed to have seen the same source in deconvolved
J-, H-, and K-band images.
In the K-band it had a relative intensity of a few percent.
The adaptive optics PSF for the Keck telescope is very
complex, and because of pupil rotation it was not possible to perform
a satisfactory deconvolution of Mkn\,231 for our data.
The feature is not included in the spectroscopic data because the slit
 orientations were such that neither overlapped with the position
 where it should be.
We are therefore unable to comment on this feature.

The H-band continuum profiles along the slit are shown in the top
panels of Fig.~\ref{fig:profiles}.
At both position angles these have been fitted with a combination of a
Gaussian and a Moffat function.
This provides an excellent match to the spatial profile because the
Gaussian represents both the core of the adaptive optics PSF and the
bright unresolved point source of the continuum related to hot dust
around the AGN; 
while the Moffat represents jointly the halo of the PSF and the faint
extended continuum component, which has wide shallow wings -- at a
radial offset of 1\arcsec, the continuum is $\sim$0.5\% of its peak
intensity.
At PA $-80^\circ$, the fit is remarkably good and has an overall FWHM
of 0.192\arcsec.
The Moffat profile has a FWHM 0.41\arcsec\ consistent with both the
expected size of the halo for the ambient seeing and the effective
radius of the de Vaucouleurs fit to imaging data \citep{tac02}.
It is similar to the
broad component at the other position angle and also those measured
in the continuum of the K-band spectra.
The Gaussian component, representing the core of the PSF, has a FWHM
0.14\arcsec.
At a PA $-25^\circ$ the Gaussian component has a slightly smaller width
giving an overall FWHM of 0.160\arcsec, consistent with a better AO
correction while these data were taken.
However, there is a residual at an offset of $-0.14$\arcsec\ which has
a peak intensity 18\% of the nucleus. 
We return to this in the discussion about the CO absorption in the
next section.

It is perhaps surprising that such a feature was not seen in the 
{\em Pueo} images of \cite{lai98}.
However, in the K-band where much of their spatial analysis was
performed, the relative strength of this feature would certainly be
much less.
Assuming the AGN component of the continuum emission is hot dust at
$\sim$1000\,K (consistent with the slopes of the H- and K-band
continua) and that the feature has a spectrum typical of late type
stars, its strength in the K-band relative to the peak would be
reduced to only 3--4\%.
In Fig.~\ref{fig:contk} we do in fact see a residual at on offset of
about $-0.2$\arcsec\ in 
spatial profile of the K-band spectrum at PA $-25^\circ$ with about
this relative strength;
but without the much clearer H-band data it would not be considered
significant.
Such a weak feature is not expected to be detected so close to the
nucleus in K-band imaging data without prior knowledge or very careful
deconvolution techniques.

\section{Nuclear Star Formation}
\label{sec:starform}

Under the assumption that the mean stellar type dominating the H-band
stellar continuum is the same across the entire region where stellar
absorption features can be seen, the absorption flux (in contrast to
the equivalent width) of the absorption features traces the spatial
extent of the stellar cluster.
We have measured this using the 1.62\micron\ CO\,6-3 bandhead
absorption (by comparing the absorption in the range
1.617--1.623\micron\ with the continuum in the range 
1.623--1.629\micron).
If there were no hot dust emission associated with the AGN, one would
expect it to be similar to the continuum profile.
However, Fig~\ref{fig:profiles} shows that the profile is indeed
substantially different to that of the continuum -- primarily in not
having a strong narrow peak in the core.
This results in a sharp reduction in the equivalent width
of the CO bandhead to $W_{\rm CO}\lesssim0.3$\AA\ at radii smaller than
0.1\arcsec;
at radii in the range 0.1--0.5\arcsec\ (beyond which the signal to
noise becomes too weak to continue), $W_{\rm CO}\sim0.7$\AA\ is
approximately constant.

\subsection{Luminosity and Mass of the Nuclear Star Formation}
\label{sec:lummass}

In late type giant and supergiant template stars \citep{mey98}, we
find that typically $W_{\rm CO}\sim4$\AA, as shown in
Table~\ref{tab:stars}.
Once such stars appear in a cluster, they will dominate
the near infrared stellar light:
in the H-band, a K2\,I star is about 10 times brighter than a O9\,I
star.
Hence, our value for $W_{\rm CO}$ implies that
dilution due to hot dust is a factor of at least 13 in the core (on
scales of 0.1\arcsec) and still a factor 5--6 around it (out to
$\sim0.5$\arcsec).
Since a partially corrected adaptive optics PSF has strong extended
emission on scales similar to the seeing, much of
the dilution on these scales may be due to the halo of the PSF and
hence attributable to the AGN rather than being associated with the
underlying and surrounding stellar population.
If this is the case, then in a 2\arcsec\ length of the slit (which includes
essentially all the near-infrared flux, and for which 
$W_{\rm CO}=0.5$\AA), only 1/8 of 
the H-band light originates in star formation.
Extrapolating to a filled circular 2\arcsec\ aperture, we estimate
that approximately 20--25\% of the H-band light is stellar.
Although the total H-band magnitude in this aperture is 10.09, the
stellar contribution is therefore 11.72\,mag.
A distance modulus of 36.15 gives an absolute magnitude for
the stellar light of M$_H=-24.43$.

We have applied no extinction correction to this value, because all
estimates of the extinction suggest that it is low:
optical and near infrared hydrogen recombination line ratios suggest
it is in the range $A_{\rm V} = 2$--6\,mag \citep{bok77,kra97},
consistent with conclusions that the inner disk of Mkn\,231 is close
to face-on and hence relatively unobscured \citep{dow98,car98}.
Instead we note that in the following discussion, applying any
extinction correction would increase the H-band luminosity, with the
result that both the mass and the bolometric luminosity of the star
forming region would increase and hence the upper limit of the
permissible age range would decrease.

Whether the stellar light is dominated by active star formation or an
old population of stars can be determined from its mass to light ratio.
Using population synthesis models for M$_V$ and V-H from Starburst99
\citep{lei99}, one can calculate the mass of stars for a given age for
both instantaneous and continuous star forming scenarios (we assume
solar metallicity and a Salpeter IMF).
Given that the stellar `cluster' (note: for convenience we use the
word `cluster' for the nuclear star forming region even though it is
not known whether the stars are distributed continuously, or if they
actually exist in many compact clusters) is extended
over about 0.40\arcsec\ (see below),
equivalent to 330\,pc, it seems unlikely that all the stars were
formed in a single instantaneous burst.
We therefore consider only continuous star forming scenarios.
The minimum possible mass of the stellar cluster occurs when
late type supergiants dominate the H-band continuum, at an age of
about 10\,Myr.
Fig.~\ref{fig:sb99} shows that to reach M$_H=-24.43$ at this age
requires a star formation rate of 
$\sim125$\msun\,yr$^{-1}$, yielding a total mass is
$1.6\times10^9$\msun\ and a light/mass ratio of 830\,$L_\odot$/\msun.
The absolute upper limit to the age is set by the fact that the
stellar mass cannot exceed the dynamical mass in the same region.
As shown in Table~\ref{tab:masses}, the
stellar mass is unlikely to exceed $5.5\times10^9$\msun.
Hence, from Fig.~\ref{fig:sb99}, the stellar cluster must be less than
120\,Myr old, over which timescale the average star formation rate
would have to be $\sim45$\msun\,yr$^{-1}$, and have a lower light/mass
ratio 125\,$L_\odot$/\msun.

\cite{car98} detected 1.4\,GHz radio continuum from a
$440\times310$\,mas disk with a flux density of 130\,mJy (see
Fig.~\ref{fig:slits}).
Further analysis of this and other 5--22\,GHz radio continuum data at
various resolutions were presented by \cite{tay99}.
Following \cite{con92}, these two sets of authors estimate the
associated star formation rate for 
stars with masses $\gtrsim5$\msun\ to be $\sim105$--115\msun\,yr$^{-1}$
(adjusted to our adopted distance for Mkn\,231).
Extending the initial mass function down to 1\msun\ will increase the
star formation rate by about a factor of two.
Alternatively, one can use Starburst 99 models to derive the star
formation rate from the supernova rate $\nu_{\rm SN}$, which the radio
continuum yields directly since the flux at 1.4\,GHz is likely to be
dominated 
by non-thermal emission (indeed the thermal continuum related to 
ionization by young stars would be only about 10\% of that measured
for the star formation rate above).
Using the Galactic relation between the non-thermal radio luminosity
and $\nu_{\rm SN}$ \citep{con92} yields $\nu_{\rm SN} = 4.3$\,yr$^{-1}$
and hence a star formation rate of order 200\msun\,yr$^{-1}$.
These estimates agree well with those above based on the near infrared
CO absorption and H-band luminosity.

One important question for Mkn\,231, which has repeatedly been
addressed through the literature, concerns the respective fractions of
the bolometric 
luminosity due to nuclear star formation and the AGN. 
Previously indirect estimates suggest that about 1/3 of the far
infrared luminosity originates in nuclear star formation:
\cite{kra97} used the 1-0\,S(1) and an analogy to the
circumnuclear ring in NGC\,7469 for which the bolometric luminosity
due to stars is known \citep{gen95};
\cite{tac02} assumed that the stars are in a disk and used the stellar
velocity dispersion and point source subtracted K-band image to derive
a mass to light ratio and hence a bolometric luminosity.
An alternative, albeit empirical, method is to use the far
infrared radio correlation with the radio continuum measurement from
\cite{car98}.
These authors used the correlation to determine whether star formation
is a reasonable explanation for the radio continuum, and derived the
ratio between the far infrared and radio fluxes to be $Q=2.5$.
The median value is $Q=2.3\pm0.2$ at 1.4\,GHz \citep{hel85}, which
would imply $\log{L_{\rm FIR}/L_\odot} = 11.92$.
This is about 70\% of the observed far infrared (40--120\micron)
luminosity of Mkn\,231, and about 25\% of the 8--1000\micron\
luminosity.

We are able to use a more direct method, based on the CO\,6-3
absorption which arises in the stars themselves.
The limits on the star formation rate and age we have discussed above
constrain the bolometric luminosity of the cluster to be 
$\log{L_{\rm bol}/L_\odot}=11.85$--12.11 (the lower luminosity
corresponding to the greater age). 
The bolometric luminosity of Mkn\,231 (taken as the luminosity in the
range 8-1000\micron) is $\log{L_{\rm bol}/L_\odot}=12.5$ 
\citep{lip94,tac02}).
Hence the star formation within 300\,pc of the AGN is responsible for
25--40\% of the galaxy's luminosity.
The range would reduce considerably if the mass of the stars could be
constrained more tightly.
For this calculation we have assumed that the star formation is
continuous.
Once it ceases, the bolometric luminosity will fall very fast, and the
cluster will fade by an order of magnitude over a timescale of up to a
few 100\,Myr.
Repeating the same calculation for smaller apertures shows that
within 0.5\arcsec\ (165\,pc, similar to the FWHM of the stellar
cluster) of the AGN, stars are still responsible for 15--30\% of
$L_{\rm bol}$;
while within a radius of 0.25\arcsec\ the fraction drops to 10--15\%
because many of the stars are then excluded.

\subsection{Spatial Distribution of the Nuclear Star Formation}
\label{sec:starprof}

\begin{figure}
\centerline{\psfig{file=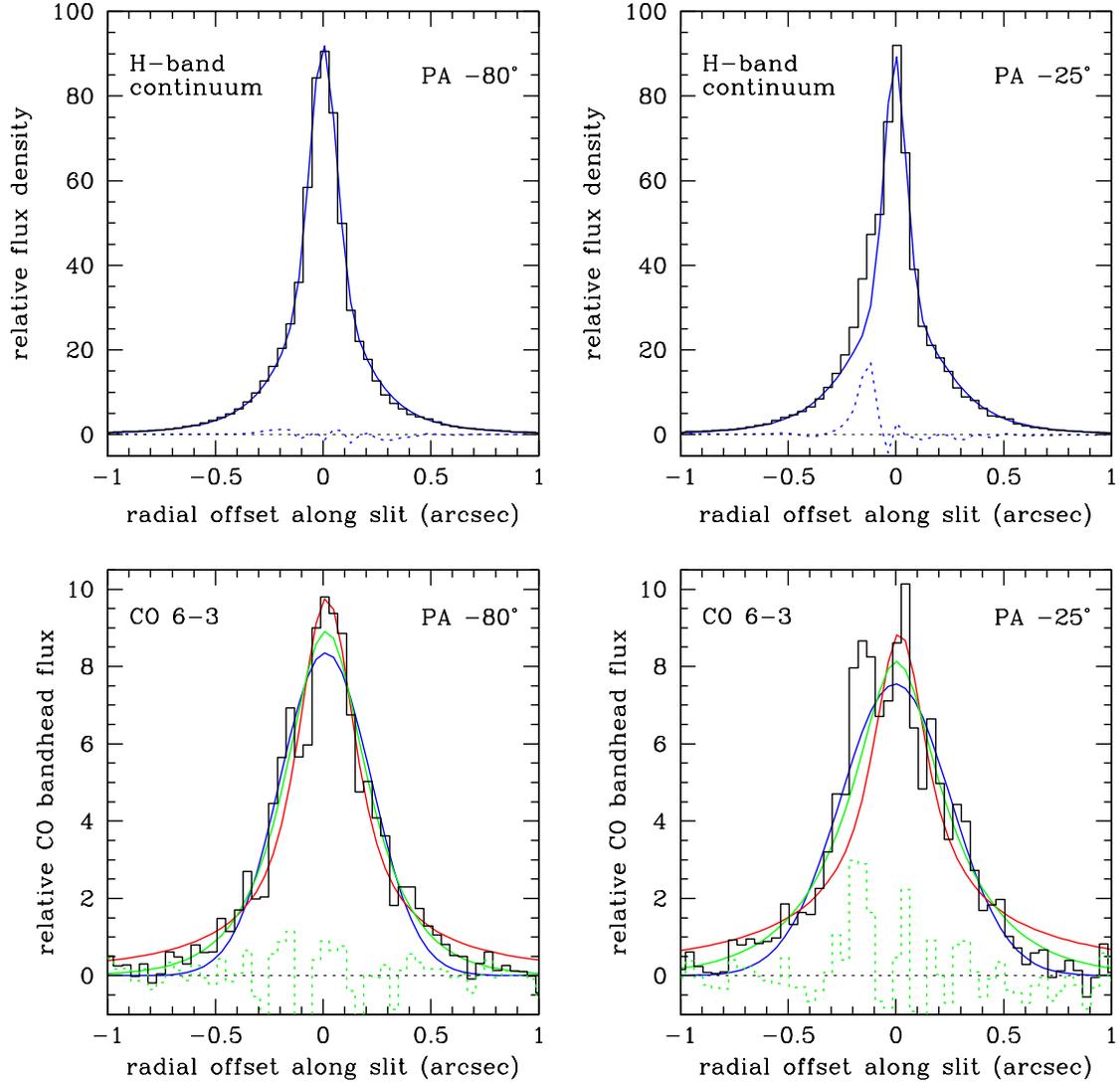,width=16cm}}
\caption{Spatial profiles of the H-band continuum (upper) and
1.62\micron\ CO\,6-3 bandhead absorption flux (lower) in the inner
2\arcsec\ of Mkn\,231, 
at PA $-80^\circ$ (left) and PA $-25^\circ$ (right).
In the upper panels, the continuum is overdrawn with a combined
Gaussian plus Moffat, which provides a very good fit to the spatial profile.
In the upper right panel, this reveals a residual at an offset of
$-0.15$\arcsec.
In the lower panels the three possible models considered in the text are
overplotted: 
blue -- a Gaussian;
red -- a de Vaucouleurs profile;
green -- an exponential profile.
The latter two are convolved (in 2 dimensions)
with a Gaussian to represent the beam.
The residual after subtracting the exponential profile is also
shown in green (dotted line)
In the lower right panel, regardless of which fit is used, the
secondary peak at $-0.15$\arcsec\ can be clearly seen.
}
\label{fig:profiles}
\end{figure}

\begin{figure}
\centerline{\psfig{file=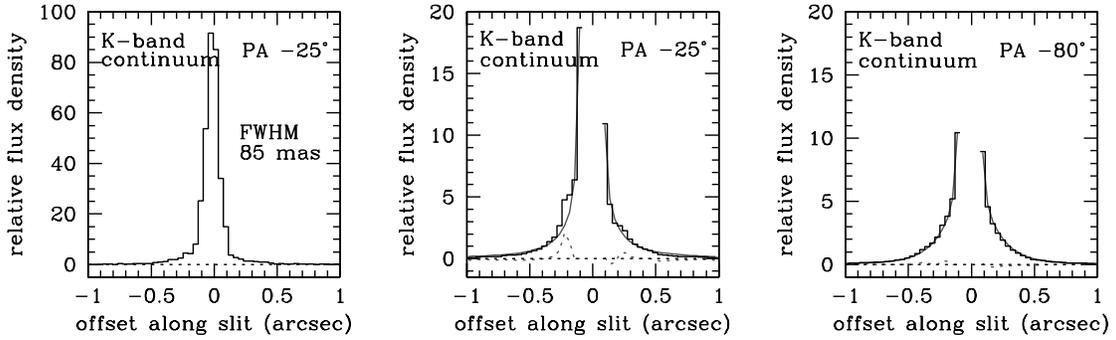,width=16cm}}
\caption{Spatial profiles of the K-band continuum for a short exposure
  (left), and long exposures (right) for the two position angles. 
All three are drawn on the same relative flux density scale.
The long exposures resulted in saturation of the core of the PSF.
A Gaussian plus Moffat profile (corresponding to the `core+halo'
  adaptive optics PSF as well as the unresolved AGN + extended stellar
  population of the galaxy) has been fitted to the measurable part of
  the profile and is shown in grey, together with its residual.
A small feature at PA $-25^\circ$ can be seen at an offset of about
  $-0.2$\arcsec\ with the expected relative strength of 2\%; 
however without the much clearer feature in the
  H-band data, this would not be considered significant.
}
\label{fig:contk}
\end{figure}

\begin{figure}
\centerline{\psfig{file=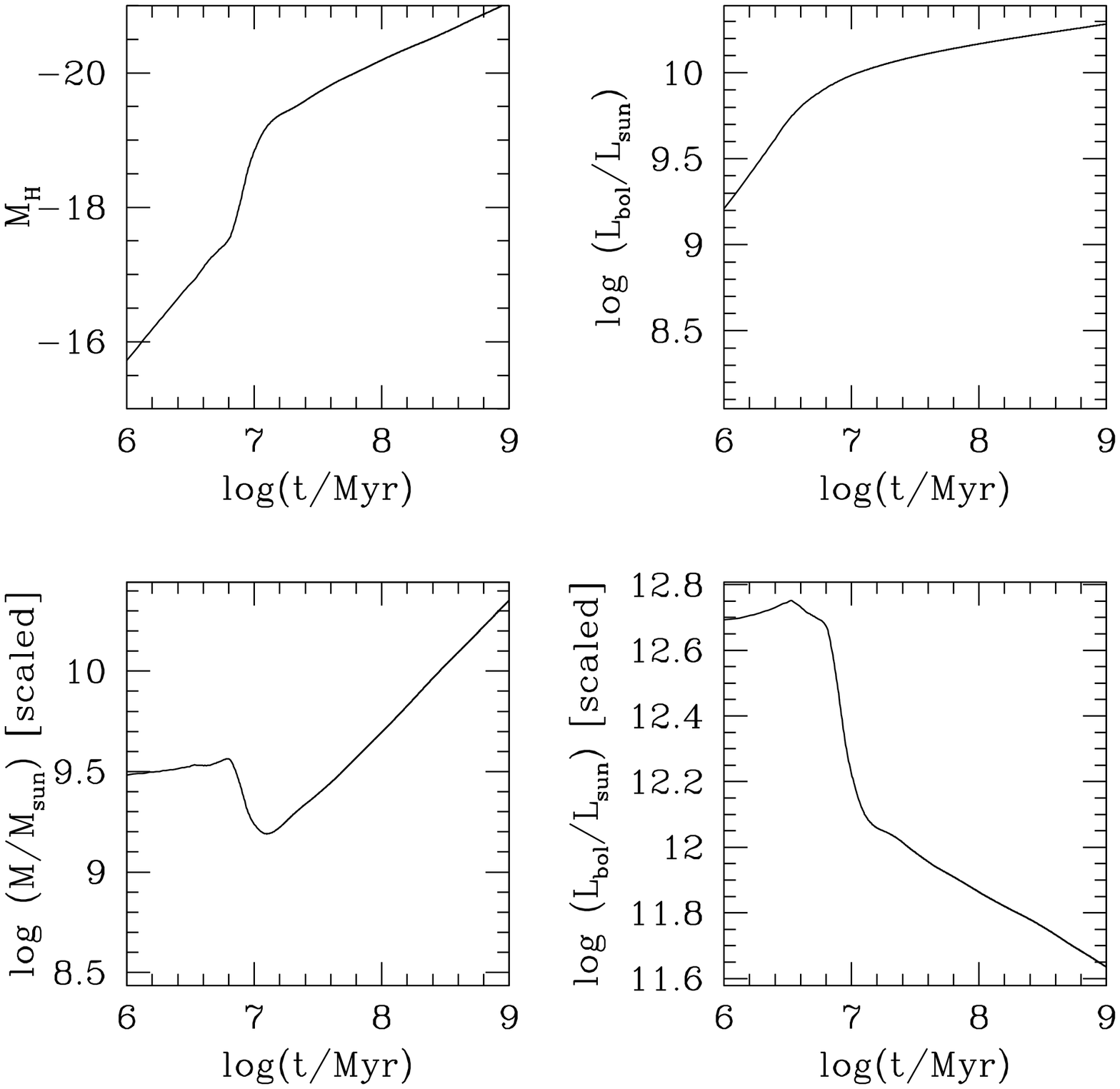,width=12cm}}
\caption{Starburst models for continuous formation of a stellar
  cluster using a solar metallicity Salpeter IMF in the range 
  1--100\msun, from {\em Starburst99} \citep{lei99}.
Top left: absolute H-band magnitude as a function of age for a star
  formation rate 1\msun/yr$^{-1}$.
Top Right: bolometric luminosity as a function of age for a star
  formation rate 1\msun/yr$^{-1}$
Bottom Left: mass of a cluster, scaled so that $M_H=-24.43$ at any given age.
Bottom Right: bolometric luminosity of a cluster, scaled so that $M_H=-24.43$
  at any given age.}
\label{fig:sb99}
\end{figure}

We have fitted three analytical functions to the CO absorption
profiles out to radii of 1.5\arcsec\ in order to assess how the
luminosity (and hence density) of the stellar cluster changes with
radius.
At PA $-25^\circ$, a small section around the residual peak seen in the
continuum was omitted in order to avoid biassing the fit.
The simplest profile is a Gaussian (blue curves in lower panels of
Fig.~\ref{fig:profiles}).
However, the best fits with FWHM 0.50\arcsec\ and 0.60\arcsec, at PAs
$-80^\circ$ and $-25^\circ$ respectively,
underestimate both the core and the wings of the measured profiles and
do not yield good matches.
It shows definitively that the spatial profile of the stellar cluster
has a sharper cusp and broad shallow wings.
We therefore consider two further profiles which have a physical
meaning when applied to stellar distributions.

The de Vaucouleurs $r^{1/4}$ profile provides an
empirical match to the luminosity profile in elliptical galaxies and
other spheroids.
The functional form 
$I(r) = I_e \exp{\{-7.67[(r/r_e)^{1/4}-1]}\}$ has been convolved (in 2
dimensions) with 
a Gaussian which represents our best estimate of the beam smearing
(FWHM 0.192\arcsec\ and 0.160\arcsec\, as described in
Section~\ref{sec:im}).
These provide acceptable fits to the inner part of the profiles, but
tend to overestimate the wings at radii beyond 0.5\arcsec\ (red curves
in Fig.~\ref{fig:profiles}).
The derived effective radii are also surprisingly large,
$r_e=1.61$\arcsec\ and 4.32\arcsec\ (1.4 and 3.5\,kpc).

The final function is an exponential profile typical of galaxy disks,
which has the form 
$I(r) = I_0 \exp{\{-r/r_d\}}$ (note that the effective radius, inside
of which half the total luminosity is emitted, is a
factor 1.68 times the disk scale length $r_d$).
This was also convolved with a Guassian representing the beam
smearing, and provides acceptable fits (green lines in the same
figure) over the whole CO profile at both position angles.
The resulting disk scale lengths are 0.18\arcsec\ and 0.24\arcsec\ (150
and 200\,pc), consistent with the size scales over which we have data.

\begin{figure}
\centerline{\psfig{file=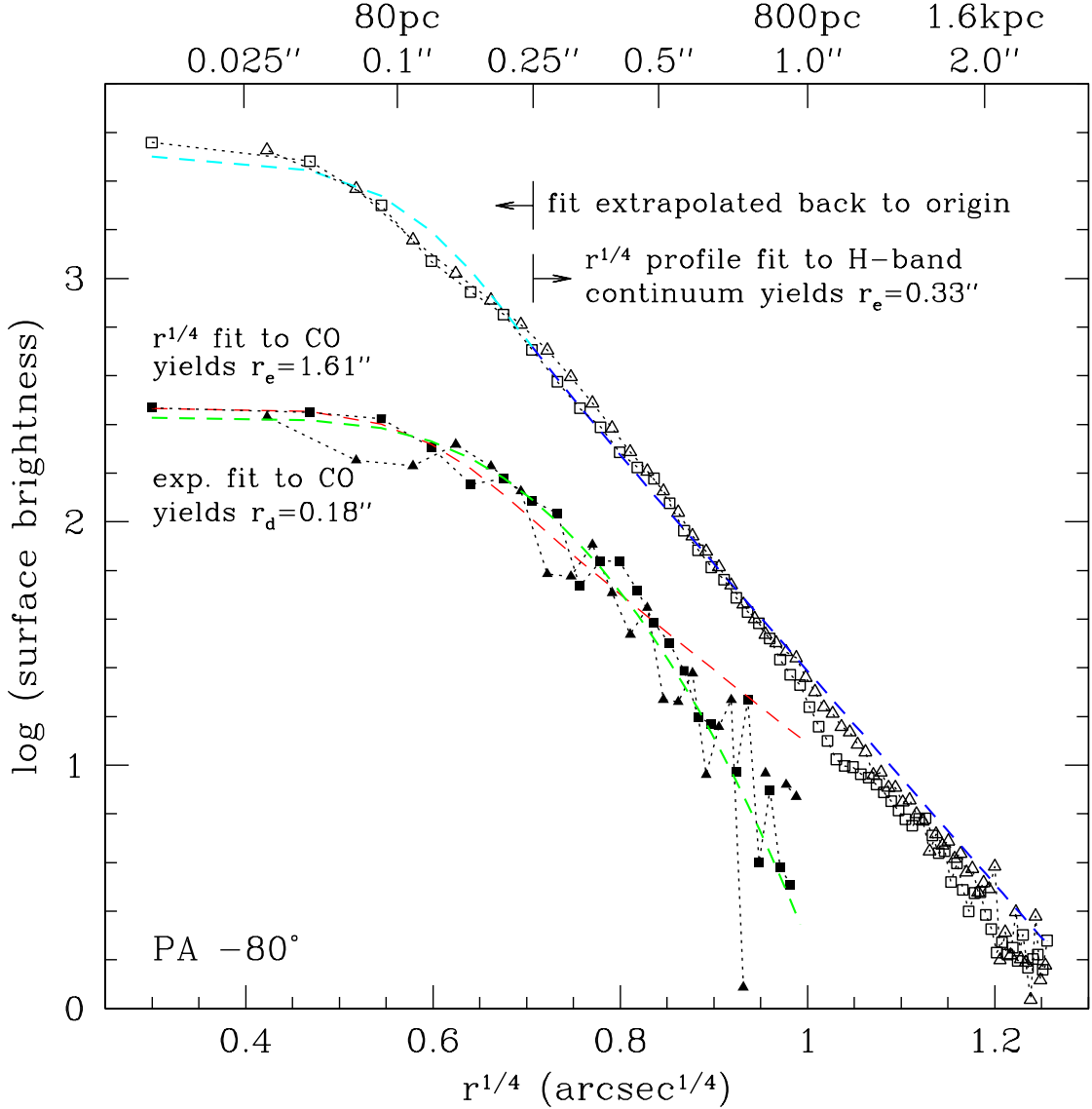,width=16cm}}
\caption{Profiles of the H-band continuum (open shapes) and CO\,6-3
  absorption flux (filled shapes) plotted as functions of $r^{1/4}$
  for PA $-80^\circ$.
The triangles and squares represent the two sides of the nucleus along
  the slit.
Contiguous points are joined by dotted lines.
Absolute scaling is arbitrary. 
The graph for PA $-25^\circ$ is similar, and is not shown.
Overplotted are (dark blue) a de Vaucouleurs fit to the continuum at
  radii 0.25--2.5\arcsec, which 
has been extrapolated back to the nucleus (cyan line).
Also shown are the de Vaucouleurs (red line) and the exponential
  (green line) fits to the CO profile at radii $<1$\arcsec.
Note that the effective radius $r_e$ define the region within which
  half the luminosity of the profile originates; the effective radius
  of the exponential profile is a facot 1.68 times larger than the disk
  scale length $r_d$ given.
}
\label{fig:dvex}
\end{figure}

Which of these two profiles best represents the data has important
implications on our understanding of the geometry of the nuclear star
forming region: in essence, whether it is spheroidal or a disk.
However, it is not immediately clear which is correct.
Previous measurements of the radial profile of the H- and K-band
continua (after subtracting a nuclear point source) yield very clear
signatures of an $r^{1/4}$ law, but with 
small effective radius $r_e=0.5$\arcsec\ and $0.25$\arcsec\ (400 and
200\,pc) respectively \citep{tac02,lai98}.
In order to reconcile this result with our fits above, we consider
also the H-band continuum profile from our spectroscopic data, which
is shown in Fig.~\ref{fig:dvex}.
To avoid contamination by the AGN component we have fit the
profile at radii 0.25--2.5\arcsec, in which region it clearly follows
the $r^{1/4}$ de Vaucouleurs function (dark blue line).
The effective radius we derive is $r_e=0.33$\arcsec\ (270\,pc),
consistent with the 0.5\arcsec\ (400\,pc) found by \cite{tac02}.
However, if we extrapolate this profile back in towards the nucleus
(convolved with our estimate of the PSF as earlier), we find that it
accounts for all of the continuum (cyan line) -- although with some
small differences to the observed profile.
Yet since we know that most of the nuclear continuum is associated with
the AGN rather than stars (e.g. from $W_{\rm CO}$), this indicates
that an $r^{1/4}$ profile with this $r_e$ cannot extend all the way in
to the nucleus.

The same figure also shows the profile of the CO absorption flux to
radii of 1\arcsec\ 
together with the $r^{1/4}$ fit (red line) and the exponential fit
(green line) from above.
It can be seen that the large $r_e=1.61$\arcsec\
for the former is needed to match the broad shape of the profile.
But such a large $r_e$ is inconsistent with that derived for the
continuum (which at large radii $>1$\arcsec\ is certainly dominated by
stellar light), since it would imply that $W_{\rm CO}$ increases rapidly
and continuously with radius -- and hence would require a
strong gradient in the stellar population.
Additionally, at radii greater than about 0.5\arcsec, the fit appears
to begin diverging from the data, although this is close to the noise
limit.
Accordingly, we are led to exclude the $r^{1/4}$ fit to the CO data.
On the other hand, the exponential fit matches the CO data well, and
does not contradict the continuum profile.

Our conclusion here is therefore that at radii greater than
$\sim$1\arcsec\ the continuum traces the luminosity profile of the
stellar population, which follows a $r^{1/4}$ profile.
Fig.~\ref{fig:dvex} hints that there may be a small break in the slope
of the profile, and that at radii larger than $\sim$0.8\arcsec\
(600-700\,pc) $r_e$ is slightly larger than the value we found above.
The result of this would be that at smaller radii, the $r^{1/4}$
profile no longer dominates the continuum.
Instead, as shown by the CO data, the stellar light here is dominated
by a bright nuclear population no more than 10--100\,Myr old, which is
present in a disk and hence has an exponential profile.
At very small radii, the continuum traces the unresolved hot
dust emission associated with the AGN.
The fact that the continuum appears to follow a single $r^{1/4}$ law
at all radii less than 2.5\arcsec\ is an unfortunate coincidence, since
it hides the true stellar luminosity profile at small radii.
A similar configuration is seen in NGC\,1068 (\citealt{tha97}, Davies
et al. in  prep.), but in that case the break between the $r^{1/4}$
profile and the nuclear exponential star forming region occurs at a
radius of $\sim$70\,pc, and is much clearer.

\subsection{An Extranuclear Star Cluster}
\label{sec:ensc}

At PA $-25^\circ$, a strong residual is apparent at an offset of
$-0.17$\arcsec\ with a peak intensity of 32\% of the central point.
The position is similar to the residual seen in the continuum.
However, this cannot be a peculiarity of the adaptive optics correction
for two reasons:
(1) the equivalent width of the CO is more than a factor 2 higher
than that at the nucleus;
(2) the residual is much narrower than the main profile; 
if a scaled replica of the main profile were to be added instead, the
residual would be far less distinct and apparent only as an asymmetry.
We conclude that this is indeed a real feature, and propose that it
may be an unresolved circumnuclear star cluster residing 120--140\,pc
from the AGN.
In fact, Fig.~\ref{fig:slits} shows that the 1.4\,GHz radio continuum
\citep{car98} also has a subsidiary peak at about this position,
providing additional evidence that it is a real feature.

We estimate that about 1\% of the CO absorption in the nuclear star
forming region originates in this feature.
If the star formation here has the same history as discussed above,
the mass would be of order $2\times10^7$\msun.
While this is a large mass for a single star cluster, it is only a
factor of a few greater than the most massive clusters seen in the
Antennae \citep{men03} and Henize\,2-10 \citep{vac02},
and is entirely consistent with the larger cluster masses found in
NGC\,6745 \citep{gri03}, Arp\,220 \citep{sha94}, and NGC\,7252
\citep{mar01}.

\section{Kinematics and Mass}
\label{sec:kin}

Mkn\,231 appears to be a later stage merger between gas rich spiral
galaxies.
In a study of 18 late stage ULIRG mergers, \cite{gen01} and
\cite{tac02} showed characteristics of relaxed stellar populations
with $r^{1/4}$ luminosity profiles and stellar dynamics dominated by
dispersion rather than rotation.
\citeauthor{tac02} showed that although Mkn\,231 itself is unusual in
having a lower than average velocity dispersion (only 115\kms),
it still has a $r^{1/4}$ luminosity profile and a low value of 
$V_{\rm rot}/\sigma=0.21\pm0.08$. 
The spheroidal mass enclosed within the effective radius of 300\,pc
would then be $6\times10^9$\msun.
And correlations of $\sigma$ with the black hole mass would yield 
$M_{\rm BH}=1.3\times10^7$\msun, which leads to a super-Eddington
AGN luminosity.
This result led these authors to consider whether the stars are
confined to the same disk that the molecular gas occupies.
In Section~\ref{sec:starform} we have shown that the nuclear stellar
population is very young and has an exponential profile.
Both of these results tend to support the conclusion that the stars
are in a disk.
However, before we continue in this vein, we need to consider whether
this is consistent with the stellar dynamics.
The disk properties determined by \cite{dow98} are an inclination
$i=10^\circ$, a rotation velocity of $V_{\rm rot}\sin{i}=60\kms$ at a
radius of 0.6\arcsec\ (500\,pc), a dynamical mass of
$12.7\times10^9$\msun\ within the same radius, and a scale height
(FWHM) of 23\,pc.
The actual rotation velocity at 500\,pc is then 345\kms.
Using the usual equation for calculating the velocity dispersion
perpendicular to the disk plane $\sigma^2 = 2 \pi G \Sigma z_0$
where $\Sigma$ is the mass surface density of the disk and $z_0$ is
the exponential scale height (for which we approximate 
$z_0 = 0.6$\,FWHM), we find a mean value of $\sigma=80$\kms.
Hence,  $V_{\rm rot}/\sigma\sim4.3$, typical of rotationally supported
systems.
That $\sigma$ is rather larger than one might expect for typical
galactic disks, for which $\sigma$ is a few times 10 to about 100\kms\ 
\citep{bot93}, is simply due to the large surface density of the disk, 
on average $1.6\times10^4$\msun\,pc$^{-2}$.
It seems therefore that it may be possible to reproduce a high
dispersion and low rotation velocity from a disk.
To test this more thoroughly we have made a model
of an exponential disk having 
$r_d=0.24$\arcsec\ and other characteristics consistent with
those above.
We have convolved it with seeing of 0.6\arcsec\ and extracted line
profiles from a 0.6\arcsec\ slit oriented at $-80^\circ$.
Integrating from $-0.7$\arcsec\ to $+0.7$\arcsec\ yields a line width
of $\sigma=107$\kms; and the range 0.3--0.7\arcsec\ yields
$V_{\rm rot}\sin{i}=35\kms$.
Thus we find an apparent $V_{\rm rot}/\sigma=0.33$, consistent with
the measurements of \cite{tac02} and showing that a nearly face-on
stellar disk can masquerade as a spheroid if it is sufficiently dense
and the details are blurred by seeing.
The caveat is that in this case, the correlation between $\sigma$ and
$M_{\rm BH}$, which is derived for dynamically hot systems, does not
apply since most of the motion is in rotation.

\subsection{Gas and Stellar Kinematics}

The gas dynamics were determined using the 2.12\micron\ H$_2$
1-0\,S(1) and the 1.64\micron\ [Fe{\sc ii}] emission lines.
The long exposures needed to detect the more extended (i.e. out to
radii of 0.5\arcsec\ or more)
line emission resulted in the central few rows of the K-band spectrum
being saturated;
due to the sub-pixel shifting associated with straightening the
spectral traces, it is not possible to use the central 6 rows.
In the H-band, the errors in the line flux and velocity across the
central few rows of the spectrum are too large for these data to be
useful.
We are therefore unable to
measure the shape of the rotation curve closer than 0.12\arcsec\
(100\,pc) to the nucleus.
We have also measured the stellar dynamics using the H-band
absorption features.
However, while these are consistent with the gas dynamics, the error
bars are too large to draw any independent conclusions.

\begin{figure}
\centerline{\psfig{file=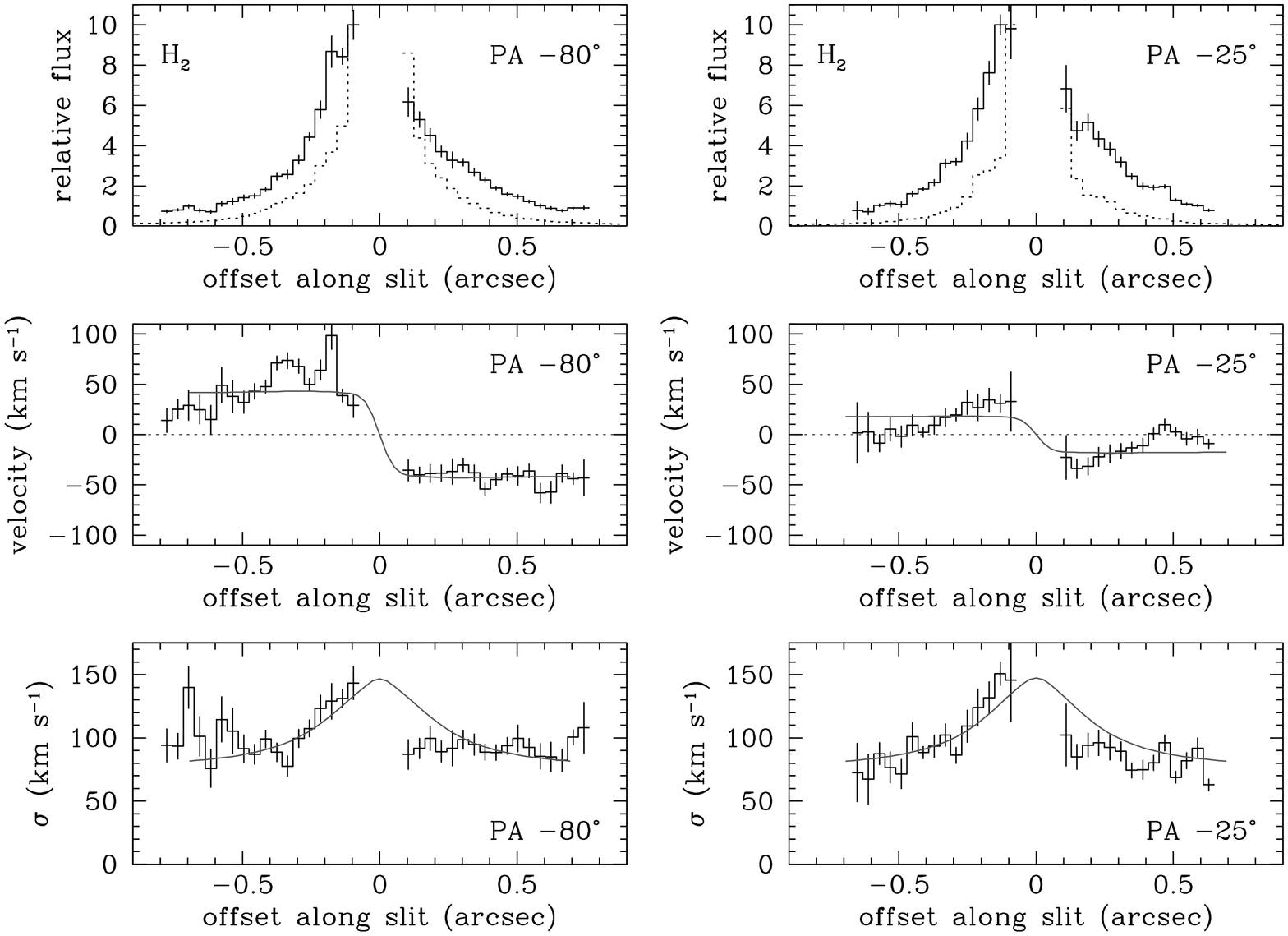,angle=-90,width=16cm}}
\caption{Spatial profiles for the 2.12\micron\ H$_2$ 1-0\,S(1) line at
  the two position angles.
Upper panel: line flux (solid lines) and continuum (dotted),
  normalised to the same maximum value; 
Centre panel: velocity curve, with respect to the systemic
  velocity of 12680\kms.
Lower panel: velocity dispersion (an approximation to the intrinsic
  dispersion can be found by quadrature correcting for the
  instrumental resolution of $\sigma=51$\kms).
The 1$\sigma$ errorbars were calculated simultaneously for all 3 parameters as
  described in the text. 
Data are shown for all rows where the spectrum was not saturated
  and the errors in the flux and velocity were above specified
  thresholds.
Overdrawn are the velocities and dispersions calculated for our simple
  mass model with $i=10^\circ$, convolved with the appropriate
  spectral and spatial beam.
}
\label{fig:h2vel}
\end{figure}

\begin{figure}
\centerline{\psfig{file=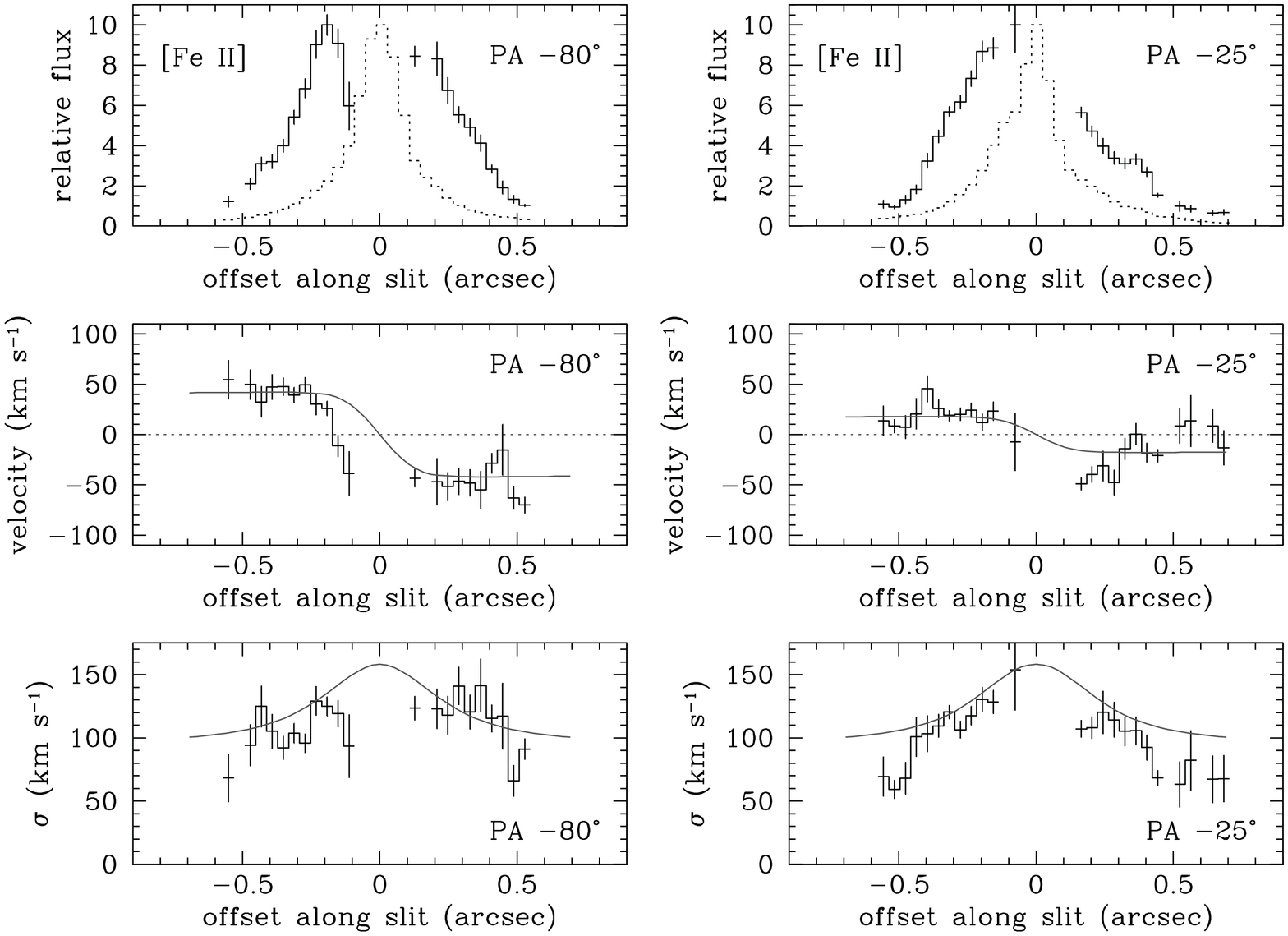,angle=-90,width=16cm}}
\caption{Spatial profiles for the 1.64\micron\ [Fe{\sc ii}] line at
  the two position angles.
Upper panel: line flux (solid lines) and continuum (dotted),
  normalised to the same maximum value; 
Centre panel: velocity curve, with respect to the systemic
  velocity of 12680\kms.
Lower panel: velocity dispersion (an approximation to the intrinsic
  dispersion can be found by quadrature correcting for the
  instrumental resolution of $\sigma=70$\kms).
The 1$\sigma$ errorbars were calculated simultaneously for all 3 parameters as
  described in the text. 
Data are shown for all rows where the spectrum was not saturated
  and the errors in the flux and velocity were above specified
  thresholds.
Overdrawn are the velocities and dispersions calculated for our simple
  mass model with $i=10^\circ$, convolved with the appropriate
  spectral and spatial beam.
}
\label{fig:fevel}
\end{figure}

The H$_2$ and [Fe{\sc ii}] emission line velocity profiles are shown
in Figs.~\ref{fig:h2vel} and~\ref{fig:fevel} respectively (centre panels),
together with the flux distribution (top) and the velocity dispersion
(bottom).
These parameters were all measured by fitting a Gaussian to the emission
line profile in each row of the spectrum.
The velocity zero points were calculated as the weighted mean
velocity, giving a value close to
12680\kms\ in each line and position angle.
Uncertainties were calculated simultaneously for all 3 parameters
by constructing an 11 pixel segment, 
which had the same statistical noise as the original spectrum (after
subtracting the 
fitted Gaussian), and back to which the fitted Gaussian profile was
added. 
A new Gaussian profile was then fitted to the segment, producing
slightly different parameters. The procedure was repeated 1000 times,
yielding the errorbars shown in the figure.

\begin{figure}
\centerline{\psfig{file=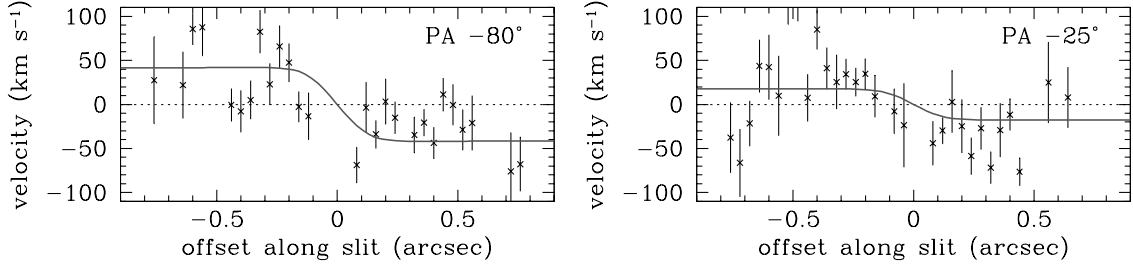,angle=-90,width=16cm}}
\caption{Velocity measurements from H-band stellar absorption
  features for PA $-80^\circ$ (left) and $-25^\circ$ (right).
Only data for which the errorbars were below a given threshold are
  included.
Overplotted in light grey is our simple mass
  model with $i=10^\circ$, convolved with the appropriate spectral and
  spatial beam.
}
\label{fig:covel}
\end{figure}

The stellar velocities were measured by cross-correlating spectra
in the range 1.57--1.64\micron\ (rest frame 1.506--1.574\micron),
extracted from each row of the dataset, with a K1.5\,I
template from \cite{mey98};
repeating the procedure with templates of different late stellar types 
gave the same results.
A Gaussian was fitted to the cross-correlation peak.
As for the emission lines, the errorbars were estimated by measuring
the residual noise in the neighbouring 51 pixels, creating a segment
with the same statistics to which the original Gaussian was added
back, and refitting the result.
While this method does not permit the velocity dispersion to be
measured, it does avoid the necessity of decovolution which can be
uncertain and adds significant noise.
The velocities and their uncertainties are plotted in
Fig.~\ref{fig:covel}.

That the cold molecular gas lies in a thin rotating disk has been
shown previously \citep{bry96,dow98,car98}.
Where we have been able to measure both the H$_2$ and [Fe{\sc ii}]
emission lines, the velocities and dispersions are remarkably
consistent and show a clear rotational signature which extends inwards
to $\sim0.1$\arcsec, as close to the nucleus as we can measure.
Thus it appears that these lines also originate in the disk
structure.
The stellar velocities are consistent with the same rotation, which
is to be expected if, as we have argued, the stars lie in the same
disk as the gas -- a scenario that
would be feasible if the stars were recently formed in the gas disk.

As a consistency check on our measurements of the emission lines, we
have calculated an 
`effective' velocity dispersion from our data in a similar aperture to
that used by \cite{tac02} who, for Mkn\,231, summed data at radii
0.3--0.7\arcsec\ either side of the nucleus.
We have first calculated the mean velocity difference from one side
of the nucleus to the other, and then combined this in quadrature with
the mean velocity dispersion in the same region.
At PA $-80^\circ$ (close to the kinematic major axis, see
Section~\ref{sec:sdm}) we find $\sigma_{\rm H_2}$ and 
$\sigma_{\rm [Fe\,II]}$ are both approximately 120\kms, consistent
with the stellar velocity dispersion reported by \citeauthor{tac02}.
At PA $-25^\circ$ (where the rotational signature is weaker) we find
$\sigma_{\rm H_2} = 65$\kms\ and  $\sigma_{\rm [Fe\,II]} = 85$\kms,
both less than the value given by these authors.
Although these results suggest that both the H$_2$ and
the [Fe{\sc ii}] lines trace the kinematic characteristics of a
rotationally supported disk in the nucleus of Mkn\,231, we consider
also whether they might be affected by shocks or outflows.

The nuclear radio continuum of Mkn\,231 has been resolved into a
triple structure oriented north-south and extended over a
distance of 0.05\arcsec\ (40\,pc) \citep{nef88,ulv99a}.
An additional component 0.1\arcsec\ (90\,pc) south of the core was
reported by \cite{tay99}.
Because the ends of these structure are elongated perpendicular to the
direction to the core, they have been interpreted as the working
surfaces of jets from the AGN.
The core itself has also been resolved on parsec scales, showing two
or three components aligned with a more east-west orientation
\citep{ulv99a,ulv99b}.
By analogy with NGC\,1068, one might expect these radio jets to be
associated with strong [Fe{\sc ii}] emission excited at the interface
between the jet and the interstellar medium \citep{bli94}.
This may be the case, but it cannot related to the [Fe{\sc ii]}
emission we have observed, which lies at much greater distances (more
than 100\,pc) from the AGN.
It therefore seems unlikely that the extended [Fe{\sc ii}] has an
origin in outflows.
On the other hand, it is co-spatial with the nuclear star forming
region, and therefore may instead originate in supernova remnants.
A star formation origin for the extremely high optical Fe{\sc ii}
luminosity of Mkn\,231 has already been proposed by \cite{lip94}.

Our confidence in this conclusion is strengthened by the similarity of
the kinematic signatures of the H$_2$ and [Fe{\sc ii}] lines which,
again by analogy with NGC\,1068, one would not necessarily expect to
be the case if they were produced in outflows rather than stellar
processes.
There are in fact only two places where the emission line kinematics
differ significantly.
One of these is the difference in velocities for small negative
offsets at PA $-80^\circ$.
The excess in both line fluxes which is also seen at an offset of
$-0.1$\arcsec\ along this position angle may be
connected to this disparate kinematic behaviour.
The second occurrence is the difference in dispersions at positive
offsets along the same position angle.
Here the errors in the dispersion for the [Fe{\sc ii}] line are
relatively large, and the difference from the H$_2$ dispersions
is therefore not highly significant.
There is only one remaining unexpected aspect.
At PA $-25^\circ$ the velocities show a symmetrical decay at
increasing radii from $\sim\pm40$\kms\ to zero within
0.3--0.4\arcsec.
This is reminiscent of Keplerian rotation about an unresolved mass.
However, the decrease in velocity is steeper than $r^{-2}$, and we
discuss it further in Section~\ref{sec:warp}.

\subsection{A Simple Dynamical Model}
\label{sec:sdm}


With only two slit positions, the inclination angle and
kinematic major axis are unconstrained by our data.
Instead, we fix them as input parameters to our model using estimates
based on interferometric measurements from the literature
of the millimetre CO emission line and radio continuum.
For the kinematic major axis, we adopt the value from
\cite{dow98}, who
used a 0.7\arcsec$\times$0.5\arcsec\ synthesised beam to measure the
properties of the CO\,2-1
line (the highest resolution data available for CO emission).
They found the kinematic line of nodes to be oriented at
$-90^\circ$ (i.e. east-west), similar to that determined by \cite{bry96} in
slightly lower resolution data. 
For the inclination angle, various estimates exist.
Based on the morphology of the nuclear radio continuum, \cite{car98}
estimated $i=45^\circ$.
However, we have argued in Section~\ref{sec:starform} that the
elongation in their map is due to additional massive star clusters
which are not intrinsically part of the nuclear disk.
Excluding these makes the remaining emission much more symmetrical,
suggesting a low inclination angle.
Such a conclusion is supported by the K-band adaptive optics image in
\cite{lai98} which has a circular appearance.
\cite{dow98} reported for their CO\,2-1 data that in order to reproduce
the spectral profile and the 
symmetrical morphology, to avoid the gas mass exceeding the dynamical
mass, and to be consistent with the lack CO absorption towards the
nucleus, the inclination had to be less than 20$^\circ$ from face-on.
Additionally, to avoid an excessively high rotation speed (i.e. rather
higher than all other galaxies in their sample), the inclination
could not be much more face-on than 10$^\circ$.
Because of the dependence of mass on $(\sin{i})^{-2}$, this
uncertainty equates to a factor of 4 in the dynamical mass.
We therefore repeat our analysis using the two limiting values of
10$^\circ$ and 20$^\circ$.

To create a model of the mass distribution and kinematics, we have
used a method outlined in \cite{tac94} that has also successfully been
applied to NGC\,7469 \citep{dav04}.
We assume that the mass is distributed in a thin disk (see
\citealt{bry96} and \citealt{dow98} for H$_2$, and \citealt{car98} for
H{\sc i}; Section~\ref{sec:starprof} for the stars); 
that the galaxy is inclined by either
10$^\circ$ or 20$^\circ$ at a position angle of 90$^\circ$;
and that the disk comprises 3 components: a nuclear stellar cluster
with exponential luminosity and mass profile as derived in
Section~\ref{sec:starform}, 
a uniform gas disk as described by
\cite{dow98} (note that these authors also describe a 3\arcsec\ outer
disk, but since we are only interested in radii $\lesssim0.7$\arcsec,
we do not include this explicitly), 
and a black hole.
A typical rotation curve that results from optimising the masses is
shown in Figs.~\ref{fig:h2vel} and~\ref{fig:fevel}.
Given that the gas dynamics could be affected by non-gravitational
motions, the model yields a reasonably good representation of the
velocities and dispersions for both lines at both position angles.

For an inclination of $i=20^\circ$ our dynamical mass is only
$1.6\times10^9$\msun\ within 0.6\arcsec.
This inclination is clearly too high since it results in a mass which
is barely the minimum we have estimated for the young stars alone
based on their luminosity, and leaves no room for the gas mass.
Including the gas mass of $1.8\times10^9$\msun\ of \citeauthor{dow98},
would suggest a maximum inclination of $15^\circ$ from face-on.
At $i=10^\circ$, the inclination preferred by \citeauthor{dow98}, we
find the enclosed mass within 
a radius of 0.6\arcsec\ is $6.7\times10^9$\msun.
That this is rather less than the mass determined by them is
due to the different rotation velocities measured at this radius (but
note that their beam size was $\sim0.6$\arcsec).
Our and their measurements differ by 20\kms, which, 
because Mkn\,231 is close to face-on, corresponds to a
factor of 1.5 in the rotation curve and hence a factor 2 in the
derived dynamical mass.
The measured velocity dispersions are reproduced by imposing a disk
thickness of 15\,pc (Gaussian FWHM), similar to that estimated by
\citeauthor{dow98}.
The increase in velocity dispersion towards smaller radii is due to
the rising mass surface density of the disk, which is on average
8000\msun\,pc$^{-2}$.
If we repeat our earlier experiment of making the same measurements as
\cite{tac02} (i.e. 0.6\arcsec\ seeing, and extracting the velocity and
dispersion over lengths of a 0.6\arcsec\ slit), we find 
$V_{\rm rot}\sin{i}=28\kms$ and $\sigma=67$\kms.
While the former is in good agreement with their rotation velocity,
the dispersion is somewhat less, indicating there there remains some
uncertainty about the disk thickness and perhaps also the mass.
The black hole mass determined by the model for $i=10^\circ$ is
$6\times10^8$\msun, similar to the mass of stars and gas within a
radius of 0.1\arcsec, indicating that this is approximately the `sphere
of influence' of the black hole (within which it dominates the
dynamics).
In our data this region is not well resolved, but if one could reach
the ~0.05\arcsec\ diffraction limit of a large telescope by better
sampling in the K-band or achieving better AO correction (e.g. with a
laser guide star) in the H-band, one could impose a much stronger
constraint on the black hole mass.

\subsection{A warp in the inner disk of Mkn\,231?}
\label{sec:warp}

\begin{figure}
\centerline{\psfig{file=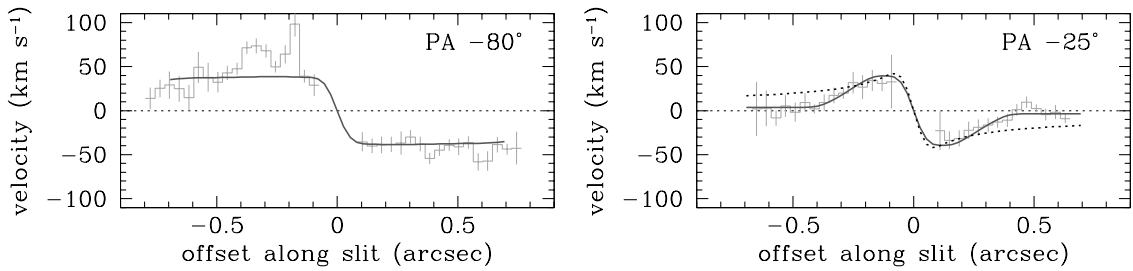,angle=-90,width=16cm}}
\caption{Velocity curves for the H$_2$ data at the
  two position angles. Overdrawn are a Keplerian $r^{-2}$ rotation
  curve (dotted line in right panel), 
and a simple warped disk model for $i=10^\circ$, showing that a
warp can explain a `faster than Keplerian' decay.
}
\label{fig:velcurve}
\end{figure}

The velocity profile at PA $-25^\circ$ shows a symmetrical decay at
increasing radii from $\sim\pm40$\kms\ to zero within
0.3--0.4\arcsec.
This is reminiscent of Keplerian rotation about a compact mass.
However, as the right panel in Fig.~\ref{fig:velcurve} indicates,
the decay is in fact faster than r$^{-2}$ and so cannot be explained
by any mass distribution confined to a planar disk.
Such a curve can in principle be produced by a warp in the inner disk of
Mkn\,231, which would have the effect of twisting the inner part of
the disk to a different inclination and position angle.
In Section~\ref{sec:starprof} we have already encountered a hint that
this may be the case:
the effective radius we derive for the exponential profile is
significantly larger at PA $-25^\circ$ than at PA $-80^\circ$,
opposite to that expected if the kinematic major axis lies at about
90$^\circ$.
Although the angular scales are small, this can be easily seen in
Fig.~\ref{fig:profiles}.
It is not an effect of possible differences in the adaptive
optics correction because the scale sizes and their difference are
much larger than those of the PSFs (in fact, both results are
independent of which PSF estimate we use).
This suggests that, at least for the stellar disk and unless it is
strongly asymmetrical, the major axis lies closer to $-25^\circ$ than
$-80^\circ$.
While it may be difficult to warp a stellar disk once it has formed,
we are proposing that the stars have formed very recently in the gas
disk.
As a result, it is certainly plausible that the warp occured in the
gas disk before the stars formed.

Even combining the kinematic and spatial data, the geometry of a
putative warp cannot be constrained.
Here, therefore, we consider only an illustrative case,
which demonstrates the feasibility of this hypothesis in explaining
the observations.
As the basic starting point, we have taken the same mass distribution
as derived in 
the previous section, and set the inclination to be $i=10^\circ$ with
the major axis at $-110^\circ$ (slightly different to the $-90^\circ$
previously adopted).
We define the position angle of the warp in the disk plane as measured
from an axis in that plane orthogonal to the major axis (i.e. if the
disk major axis runs east-west then, if the disk is deprojected, the
position angle of the warp is measured east of north).
At a position angle of 30$^\circ$ and for radii $<0.2$\arcsec, we tilt
the disk out of its plane by 10$^\circ$.
Beyond this, the tilt declines uniformly with increasing 
radius, until it reaches zero for radii $>0.35$\arcsec.
The observable effect this slight warping of a nearly face-on disk is
in fact to shift the major axis of 
the inner part of the disk more towards $-25^\circ$ without greatly
altering the effective inclination.
For this example, the inner part of the disk would have an
effective inclination of $10^\circ$ but at a position angle of
$-50^\circ$.
Hence, as shown in the right hand panel of Fig.~\ref{fig:velcurve},
at PA $-25^\circ$, the signature of rotation is seen at small
radii but not at large radii.
Comparison to Fig.~\ref{fig:covel} suggests that the stellar velocity
curve might also match this model better, again a result which one
might expect if the stars have formed recently in a disk which was
already warped.
Finally, our warp model is also able to reproduce approximately the
velocity contours presented by \cite{dow98} with a 0.6\arcsec\ beam.

We have provided an example of a weak warp which can explain the
kinematics we have observed.
However, measurements of both hydroxyl and the radio continuum provide
evidence that there may be much 
stronger warping occuring in the nucleus of Mkn\,231.
The velocity field of the 1.6\,GHz OH line emission, observed at a
resolution of 30--40\,mas over an extent of 100--150\,mas by
\cite{klo03}, has been modelled in terms of a thick disk or torus
inclined by 56$^\circ$ at a position angle of 35$^\circ$.
This orientation, on scales of 100\,pc, is rather different from that
on scales of $\sim500$\,pc traced by the molecular gas, and is
indicative of a warp completely consistent with -- but more pronounced
than -- that described above.
As we noted earlier, radio continuum data suggest that there is a very
pronounced shift in the direction of the radio jet at radial
scales between 1 and 20\,pc \citep{ulv99a,ulv99b}.
Unless the jets are viewed nearly end-on, this represents a
substantial mislignment of the jet axis.
Whether the misalignment is due to deflection of the jets as it
strikes dense interstellar medium, or from a twisting of the axis of
the central AGN, it could plausibly be related (whether causally or
not) to the warping of the nuclear disk.
Two such possibilities are that
(1) the jet impinges on the disk and has caused it to warp, at the
same time changing the direction of the jet;
(2) the spin axis in the core has changed, perhaps during the later
stages of the merger as clumps of gas thrown off in the early stages
are accreted, 
and this is reflected in both the altered jet direction and the
kinematics of the disk.
The consequence of a warp is to make modelling of the small scale
kinematics rather more complex.
However, with the right data it may also make it easier to measure the
dynamical effect of the black hole because for
such a low inclination of the disk, small uncertainties in
either the measured kinematics or the inclination itself propagate
into very large uncertainties in the true rotation curve and hence
enclosed mass estimate.
A warp will almost inevitably lead to the inner region having a larger
inclination, and hence reduce the sensitivity of the derived rotation
curve to uncertainties in these parameters.
With the current data we are unable to constrain the geometry of the
warp and hence are unable to make detailed statements about the black
hole mass.
It is expected that with either integral field spectroscopy at
adaptive optics scales, or higher resolution interferometric data at
mm wavelengths will help to resolve the situation.

\section{A Young Starburst in Mkn\,231}
\label{sec:young}

In the previous Sections we have presented evidence that there is a
massive (log $L_{bol}/L_\odot \sim 12.0$) young ($\lesssim120$\,Myr)
starburst occuring in the nucleus of Mkn\,231.
The stars involved in this burst are distributed in the same disk
plane occupied by the gas, which is nearly face-on.
This evidence suggests that the H and K band stellar light is
not tracing stars from the progenitor galaxies which have relaxed
into a spheroid during the merger, but instead is dominated by stars
which have formed {\em in situ} in the gas disk -- which itself has
formed as a result of the merger -- and still remain there.

It is important, therefore, to consider whether other observations are
consistent with such a scenario.
As we mentioned in Section~\ref{sec:lummass}, the radio continuum
supports the existence of a young active starburst.
Here we also look at the important characteristics observed with the
{\em Infrared Space Observatory}, in particular the mid infrared
emission lines reported in \cite{gen98} and the polycyclic aromatic
hydrocarbon (PAH) features in \cite{rig99}.

\cite{rig99} presented measurements of the 7.7\micron\ PAH feature for
62 ULIRGs, comparing them templates from 23 AGN and 15 starburst
and normal galaxies.
They showed that AGN and starbursts could be distinguished by the
ratio of the PAH strength at 7.7\micron\ to the continuum at
5.9\micron\ ($L/C$): AGN have $\langle L/C \rangle =0.4$, while
starbursts have $\langle L/C \rangle =2.77$.
Under this classification, Mkn\,231 with $L/C=0.3$, is clearly an AGN.
However, this does not mean that there is no starburst.
The spectrum in Fig.~2 of \cite{rig99} clearly shows a strong PAH
feature -- the low $L/C$ ratio arises simply because the continuum is
also strong.
Indeed, comparison with the other `starburst-like' ULIRGs in this
study shows that the absolute stength of the PAH feature is consistent
with what one would expect from a massive starburst.

\cite{gen98} did not detect 12.8\micron\ [Ne{\sc ii}], a line that
arises in H{\sc ii} regions and commonly seen in the mid infrared
spectra of starburst and ULIRGs.
Ne$^+$ has an ionization potential of 21.6\,eV, and so is a tracer of
the hottest stars.
The limit of $1.6\times10^{-16}$\,W\,m$^{-2}$ is a little lower than
the flux that might be expected, but on the other hand, one cannot
rule out that the true line flux lies within the range of typical
values for massive young starbursts.
A line that was detected is [Si{\sc ii}] at 34\micron.
The ionization potential for Si$^+$ is only 8.2\,eV and hence this
line arises in H{\sc ii} regions, at the interface between the ionized
and molecular gas.
The flux measured was $\sim5\times10^{-16}$\,W\,m$^{-2}$, again
similar to the other ULIRGs.
All the ULIRGs in this sample had similar fluxes for the two lines;
and the detection of [Si{\sc ii}] but non-detection of [Ne{\sc ii}]
at 1/3 the flux level is not an implausible result, particularly given
the difficulty of detecting the lines as a result of the low line to
continuum ratio.
A re-analysis of the data by A.~Verma (private communication) suggests
that there is a possible $\sim3\sigma$ detection of [Ne{\sc ii}].

One final line of evidence about this is the near to far infrared
spectral energy distribution (SED).
This has been modelled in detail by \cite{ver99}, who found that a
combination of nearly face-on torus model together with a zero age
starburst was needed to match the 1-1000\micron\ SED.
Although the exact parameters are not uniquely constrained, this
general conclusion was inescapable.

We conclude that the radio continuum, mid infrared diagnostics, and
SED all support the view that there is vigorous young star formation
occuring in the nucleus of Mkn\,231.
The superposition of this starburst and the AGN have made Mkn\,231
look similar to a quasar, but at the same time means that it does not
fit any of the standard correlations.
Once the star formation ceases, and the starburst fades -- on
timescales of one to a few 100\,Myr -- the galaxy will probably look
more like a typical Seyfert~1.

Evidence for the existence of young nuclear stellar disks in a number
of spiral galaxies has been presented in the literature in terms of a
drop in the stellar velocity dispersion on small spatial scales.
\cite{mar03} measured the $\sim$8600\AA\ Ca{\sc ii} triplet in 5 isolated
Seyfert spiral galaxies and found a dip in the velocity
dispersion across the central 1--3\arcsec\ for 4 of them (and a hint
of such in the fifth).
The drop was accompained by a possible local increase in the
equivalent width of the  Ca{\sc ii} triplet, indicating the presence
of a younger stellar population.
A similar effect has been previously reported for 3 barred galaxies by
\cite{ems01} using the 2.29\micron\ CO bandhead.
This phenomenon has been modelled by \cite{woz03} using simulations
which include stars, gas, and star formation.
Their conclusion was that the drop in the line-of-sight stellar velocity
dispersion is the kinematical signature of stars formed in a
dynamically cold gas disk, which forms as gas is driven to the nucleus
by the bar.
Because these stars are young, they out-shine the old (dynamically
hotter) population in the nucleus.
And the effect is enhanced because the high gas density in the very
centre leads to more efficient cooling, reducing further the velocity
dispersion of the gas itself, and hence also the stars formed from
this gas.
Thus, although the frequency of such nuclear disks is unkown, it may
be a relatively common phenomenon in galaxies where gas is driven to
the nucleus -- whether this be by a bar in a spiral galaxy as for the
examples above, or due to a merger event as we have seen for Mkn\,231.

\section{Conclusions}
\label{sec:conc}

We have presented adaptive optics spectroscopic data of the nearby
ultraluminous infrared galaxy and QSO/Seyfert\,1 Mkn\,231 in the H-
and K-bands at resolutions as small as 0.085\arcsec.

\begin{enumerate}
\item
We have resolved the region of active star formation in the nucleus of
Mkn\,231 through the 1.62\micron\ CO absorption in late type stars.
The region has a FWHM of 0.35--0.40\arcsec\ (300\,pc), and has a
spatial profile significantly more cusped and with broader wings than
a Gaussian.
It is fit well by an exponential profile with
disk scale length $r_d=0.18$--0.24\arcsec.
At larger radii the continuum luminosity profile is
well represented by an $r^{1/4}$ law, as one might expect, since during
a merger the stars relax into a spheroid; 
but that this also appears to
continue inwards is an unfortunate coincidence.
The nuclear star forming region would have a minimum mass of
$1.6\times10^9$\msun\ if it is only $\sim10$\,Myr old (i.e. it has
just reached the age when late
type supergiants dominate the near infrared stellar continuum),
implying a star formation rate of $\sim125$\msun\,yr$^{-1}$.
Constraints from the dynamical mass indicate that its age is unlikely
to exceed 120\,Myr.
It is responsible for 25--40\% of the galaxy's entire bolometric
luminosity, the higher fraction corresponding to the younger age
limit.
These results all point to the fact that the stars lie in a disky rather
than spheroidal distribution, and that they are very young.
Hence it is likely that they have formed recently in the nearly
face-on molecular gas disk, which is itself a product of the merger
that resulted in Mkn\,231.

\item
There is an unresolved extranuclear star cluster $\sim0.15$\arcsec\
(120--140\,pc) from the nucleus, with a mass of $\sim2\times10^7$\msun.
This is more massive than `typical' young star clusters in star
forming galaxies, but is comparable with the most massive clusters
seen in a few cases.

\item
The dynamics at radii 0.1--0.7\arcsec\ have been traced using
the 2.12\micron\ 1-0\,S(1) H$_2$ and 1.64\micron\ [Fe{\sc ii}]
emission lines, as well as the H-band stellar absorption features.
The emission lines are extended over the same spatial scales as the
stellar CO absorption, which is much greater than the scale of the
radio jets, and are therefore likely to originate in stellar processes.
Their kinematics are similar, and in general terms consistent with
rotation of a nearly face-on disk.
However, a simple dynamical model shows that there are distinct
differences.
The rotation curve at a position angle of $-25^\circ$ shows a rapid
decrease in velocity with radius, faster than $r^{-2}$.
This can be explained if there is a warp in the inner 0.2--0.3\arcsec\
(200\,pc) of Mkn\,231, a hypothesis which appears to be supported by
the orientation of the major axis of the stellar disk, by the
pronounced shift on small scales of the radio jet axis, and by the
velocity field of the $\sim150$\,mas 1.6\,GHz OH line emission.
Integral field velocity maps at 0.1\arcsec\ resolution will
be needed to constrain the geometry and properties of such a warp.

\end{enumerate}


\acknowledgments

The authors are grateful to the support of the staff of the Keck
Observatory. 
The authors wish to recognize and acknowledge the very significant
cultural role and reverence that the summit of Mauna Kea has 
always had within the indigenous Hawaiian community.  We are most
fortunate to have the opportunity to conduct observations
from this mountain.
We are grateful to A.~Verma for looking again at the ISO data of
Mkn\,231, and for making her work on its SED available to us.



\clearpage


\begin{deluxetable}{llrrrr}

\tablewidth{0pt}
\tabletypesize{\small}
\tablecaption{Late type stellar parameters\label{tab:stars}}
\tablehead{

\colhead{star} & 
\colhead{stellar} & 
\colhead{$M_{\rm V}$} & 
\colhead{$M_{\rm H}$} & 
\colhead{Mass} & 
\colhead{$W_{\rm CO}$} \\

\colhead{} & 
\colhead{type} & 
\colhead{} & 
\colhead{} & 
\colhead{($M_\odot$)} & 
\colhead{(\AA)} 

}
\startdata

HR 7479   & G4 I     & $-6.2$ & $-7.9$ & 12\phm{.0} & 1.3 \\
HR 8465   & K1.5 I   & $-5.9$ & $-8.1$ & 13\phm{.0} & 4.5 \\
HR 1155   & M2 I     & $-5.6$ & $-9.5$ & 19\phm{.0} & 5.8 \\
HR 7328   & G9 III   & $+0.8$ & $-1.3$ & 1.1        & 2.1 \\
HR 8317   & K1 III   & $+0.6$ & $-1.8$ & 1.1        & 2.7 \\
HR 165    & K3 III   & $+0.3$ & $-2.6$ & 1.2        & 3.3 \\
HR 4517   & M1 III   & $-0.5$ & $-4.3$ & 1.2        & 4.2 \\

\enddata

\tablecomments{Cols 1--2 give the names and types of the stellar
templates from \cite{mey98}, 
and col 6 is the 1.62\,$\mu$m CO\,6-3 equivalent width measured from these;
data in cols 3--5 are from \cite{cox00}.}

\end{deluxetable}


\begin{deluxetable}{lll}
\tablewidth{0pt}
\tabletypesize{\small}
\tablecaption{Derived Masses \label{tab:masses}}
\tablehead{

\colhead{Description} & 
\colhead{Mass (\msun)} & 
\colhead{Comments} \\ 

\colhead{} & 
\colhead{$r<0.6$\arcsec/500\,pc} & 
\colhead{} 

}
\startdata

Young Stars   
  & $>1.3\times10^9$\ \tablenotemark{a} 
  & minimum mass possible for $M_H=-24.43$ \\
Molecular Gas 
  & $\phm{>,}1.8\times10^9$ 
  & from CO model of \cite{dow98} \\
Dynamical\tablenotemark{b}, $i=20^\circ$ 
  & $\phm{>,}1.6\times10^9$ 
  & inconsistent with stellar and gas mass \\
Dynamical\tablenotemark{b}, $i=10^\circ$ 
  & $\phm{>,}6.7\times10^9$\ \tablenotemark{c} 
  & implies total (old+young) stellar mass $4.3\times10^9$\msun\tablenotemark{a} \\

\enddata

\tablenotetext{a}{\ Stellar masses are 25\% higher than given here if
  integrated out to $r=1.0$\arcsec\ as used in
  Section~\ref{sec:lummass}. They are given here to 
  $r=0.6$\arcsec\ for consistency with other masses.}
\tablenotetext{b}{\ Derived in Section~\ref{sec:sdm} from axisymmetric
  disk model of H$_2$ and [Fe{\sc ii}] rotation and velocity dispersion.}
\tablenotetext{c}{\ The dynamical mass derived by \cite{dow98} is a
  factor 2 higher due to their higher measured rotation velocity at
  0.6\arcsec.}

\end{deluxetable}


\clearpage


\end{document}